\documentclass[conference]{IEEEtran}

\usepackage{graphicx}
\usepackage{comment}
\usepackage{float}
\usepackage{makecell}
\usepackage{array}

\usepackage{hyperref}
\usepackage{cite}
\usepackage{amsmath}
\usepackage{amsmath,amssymb,amsfonts}
\usepackage{textcomp}
\usepackage{xcolor}
\usepackage{lscape}
\usepackage{pdflscape}

\usepackage{geometry}
\usepackage{booktabs}
\usepackage{wasysym}
\usepackage{multirow}
\usepackage[inline]{enumitem}
\usepackage{csquotes}
\usepackage{tikz}
\usetikzlibrary{turtle}
\usepackage{acro}
\usepackage[english]{babel}
\usepackage{tabularx}
\usepackage{longtable}
\usepackage{colortbl}
\usepackage{array}
\usepackage{makecell}
\usepackage[normalem]{ulem}
\useunder{\uline}{\ul}{}
\usepackage{enumitem}
\usepackage{colortbl}
\usepackage{graphicx} 

\usepackage{algorithm}
\usepackage{algorithmicx}
\usepackage{algpseudocode}
\usepackage[table]{xcolor}

\usepackage{pifont}

\usepackage[utf8]{inputenc}

\usepackage{placeins}

\usepackage{listings}

\lstdefinestyle{mystyle}{
    backgroundcolor=\color{gray!10},   
    commentstyle=\color{gray},
    keywordstyle=\color{blue},
    stringstyle=\color{red},
    basicstyle=\ttfamily\footnotesize,
    breakatwhitespace=false,         
    breaklines=true,                 
    captionpos=b,                    
    keepspaces=true,                 
    numbers=left,                    
    numbersep=5pt,                  
    showspaces=false,                
    showstringspaces=false,
    showtabs=false,                  
    tabsize=2
}

\newcommand*\emptycirc[1][1ex]{\tikz\draw (0,0) circle (#1);} 

\newcommand*\fullcirc[1][1ex]{\tikz\fill (0,0) circle (#1);}

\newcommand{\smallfullcirc}{\raisebox{-0.1em}{\scalebox{0.75}{\fullcirc}}}
\newcommand{\smallemptycirc}{\raisebox{-0.1em}{\scalebox{0.75}{\emptycirc}}}

\definecolor{r0}{rgb}{1, 1.00, 1.00}
\definecolor{r1}{rgb}{1, 0.90, 0.90}
\definecolor{r2}{rgb}{1, 0.80, 0.80}
\definecolor{r3}{rgb}{1, 0.70, 0.70}
\definecolor{r4}{rgb}{1, 0.60, 0.60}
\definecolor{r5}{rgb}{1, 0.50, 0.50}
\definecolor{r6}{rgb}{1, 0.40, 0.40}
\definecolor{r7}{rgb}{1, 0.30, 0.30}
\definecolor{r8}{rgb}{1, 0.20, 0.20}
\definecolor{r9}{rgb}{1, 0.10, 0.10}
\definecolor{r10}{rgb}{1, 0.00, 0.00}
\definecolor{g0}{rgb}{1.00, 1, 1.00}
\definecolor{g1}{rgb}{0.90, 1, 0.90}
\definecolor{g2}{rgb}{0.80, 1, 0.80}
\definecolor{g3}{rgb}{0.70, 1, 0.70}
\definecolor{g4}{rgb}{0.60, 1, 0.60}
\definecolor{g5}{rgb}{0.50, 1, 0.50}
\definecolor{g6}{rgb}{0.40, 1, 0.40}
\definecolor{g7}{rgb}{0.30, 1, 0.30}
\definecolor{g8}{rgb}{0.20, 1, 0.20}
\definecolor{g9}{rgb}{0.10, 1, 0.10}
\definecolor{g10}{rgb}{0.00, 1, 0.00}

\definecolor{o0}{rgb}{1.00, 0.90, 0.70}  
\definecolor{o1}{rgb}{1.00, 0.85, 0.55}
\definecolor{o2}{rgb}{1.00, 0.75, 0.40}
\definecolor{o3}{rgb}{1.00, 0.65, 0.25}
\definecolor{o4}{rgb}{1.00, 0.55, 0.10}
\definecolor{o5}{rgb}{1.00, 0.50, 0.00}  
\definecolor{o6}{rgb}{0.90, 0.45, 0.00}
\definecolor{o7}{rgb}{0.80, 0.40, 0.00}
\definecolor{o8}{rgb}{0.70, 0.35, 0.00}
\definecolor{o9}{rgb}{0.60, 0.30, 0.00}
\definecolor{o10}{rgb}{0.50, 0.25, 0.00} 

\newcommand{\pct}[1]{%
  \ifdim #1pt > 30pt
    \textcolor{green!50!black}{#1}%
  \else
    \ifdim #1pt < 10pt
      \textcolor{red!70!black}{#1}%
    \else
      \textcolor{orange!90!black}{#1}%
    \fi
  \fi
}

\lstdefinelanguage{JavaScriptX}{
  morekeywords={
    function, return, import, from, const, let, var, class,
    export, default, async, await, if, else, for, while, switch, case,
    new, this, typeof, try, catch
  },
  sensitive=true,
  morecomment=[l]{//},
  morecomment=[s]{/*}{*/},
  morestring=[b]',
  morestring=[b]",
  alsoletter={.},
}

\usepackage{booktabs}
\usepackage{makecell}
\usepackage{caption}

\DeclareAcronym{ai}{
  short = AI,
  long  = artificial intelligence
}

\DeclareAcronym{rl}{
  short = RL,
  long  = reinforcement learning
}

\DeclareAcronym{icq}{
  short = ICQ,
  long  = Interchain Queries
}

\DeclareAcronym{dex}{
  short        = DEX,
  short-plural = DEXs,
  long         = decentralised exchange,
  long-plural  = decentralised exchanges
}

\DeclareAcronym{hrmp}{
  short = HRMP,
  long  = Horizontal Relay-Routed Message Passing
}

\DeclareAcronym{xcm}{
  short = XCM,
  long  = Cross-Consensus Messaging
}

\DeclareAcronym{api}{
  short = API,
  long  = Application Programming Interface
}

\DeclareAcronym{eoa}{
  short = EOAs,
  long  = Externally Owned Accounts
}

\DeclareAcronym{sca}{
  short = SCAs,
  long  = Smart Contract Accounts
}

\DeclareAcronym{dapps}{
  short = dApps,
  long  = decentralised applications
}

\DeclareAcronym{dapp}{
  short = dApp,
  long  = decentralised application
}

\DeclareAcronym{defi}{
  short = DeFi,
  long  = decentralised finance
}

\DeclareAcronym{dlt}{
  short = DLT,
  long  = distributed ledger technology
}

\DeclareAcronym{tvl}{
  short = TVL,
  long  = total value locked
}

\DeclareAcronym{dl}{
  short = DL,
  long  = deep learning
}

\DeclareAcronym{fl}{
  short = FL,
  long  = federated learning
}

\DeclareAcronym{zkml}{
  short = ZKML,
  long  = zero-knowledge machine learning
}

\DeclareAcronym{iot}{
  short = IOT,
  long  = internet of things
}

\DeclareAcronym{iiot}{
  short = IIoT,
  long  = industrial internet of things
}

\DeclareAcronym{amm}{
  short        = AMM,
  short-plural = AMMs,
  long         = automated market maker,
  long-plural  = automated market makers
}

\DeclareAcronym{aigc}{
  short = AIGC,
  long  = AI-generated content
}

\DeclareAcronym{did}{
  short        = DID,
  short-plural = DIDs,
  long         = decentralised identifier,
  long-plural  = decentralised identifiers
}

\DeclareAcronym{pbft}{
  short = PBFT,
  long  = Practical Byzantine Fault Tolerance
}

\DeclareAcronym{dag}{
  short = DAG,
  long  = directed acyclic graph
}

\DeclareAcronym{vrf}{
  short        = VRF,
  short-plural = VRFs,
  long         = verifiable random function,
  long-plural  = verifiable random functions
}

\DeclareAcronym{evm}{
  short = EVM,
  long  = Ethereum Virtual Machine
}

\DeclareAcronym{wasm}{
  short = WASM,
  long  = WebAssembly
}

\DeclareAcronym{tee}{
  short        = TEE,
  short-plural = TEEs,
  long         = trusted execution environment,
  long-plural  = trusted execution environments
}

\DeclareAcronym{ml}{
  short = ML,
  long  = machine learning
}

\DeclareAcronym{llm}{
  short        = LLM,
  short-plural = LLMs,
  long         = large language model,
  long-plural  = large language models
}

\DeclareAcronym{p2p}{
  short = P2P,
  long  = peer-to-peer
}

\DeclareAcronym{cnn}{
  short        = CNN,
  short-plural = CNNs,
  long         = convolutional neural network,
  long-plural  = convolutional neural networks
}

\DeclareAcronym{lstm}{
  short        = LSTM,
  short-plural = LSTMs,
  long         = long short-term memory,
  long-plural  = long short-term memories
}

\DeclareAcronym{gnn}{
  short        = GNN,
  short-plural = GNNs,
  long         = graph neural network,
  long-plural  = graph neural networks
}

\DeclareAcronym{ddpg}{
  short = DDPG,
  long  = deep deterministic policy gradient
}

\DeclareAcronym{drl}{
  short = DRL,
  long  = deep reinforcement learning
}

\DeclareAcronym{gan}{
  short        = GAN,
  short-plural = GANs,
  long         = generative adversarial network,
  long-plural  = generative adversarial networks
}

\DeclareAcronym{nlp}{
  short = NLP,
  long  = natural language processing
}

\DeclareAcronym{xai}{
  short = XAI,
  long  = explainable artificial intelligence
}

\DeclareAcronym{powu}{
  short = PoUW,
  long  = proof of useful work
}

\DeclareAcronym{pole}{
  short = PoLe,
  long  = proof of learning
}

\DeclareAcronym{tps}{
  short = TPS,
  long  = transactions per second
}

\DeclareAcronym{nft}{
  short        = NFT,
  short-plural = NFTs,
  long         = non-fungible token,
  long-plural  = non-fungible tokens
}

\DeclareAcronym{poc}{
  short = PoC,
  long  = proof of contribution
}

\DeclareAcronym{poi}{
  short = PoI,
  long  = proof of interpretability
}

\DeclareAcronym{potc}{
  short = PoTC,
  long  = proof of trust collaboration
}

\DeclareAcronym{pot}{
  short = PoT,
  long  = proof of thought
}

\DeclareAcronym{dp}{
  short = DP,
  long  = differential privacy
}

\DeclareAcronym{smpc}{
  short = SMPC,
  long  = secure multi-party computation
}

\DeclareAcronym{he}{
  short = HE,
  long  = homomorphic encryption
}

\DeclareAcronym{zkp}{
  short        = ZKP,
  short-plural = ZKPs,
  long         = zero-knowledge proof,
  long-plural  = zero-knowledge proofs
}

\DeclareAcronym{ipfs}{
  short = IPFS,
  long  = InterPlanetary File System
}

\DeclareAcronym{vc}{
  short        = VC,
  short-plural = VCs,
  long         = verifiable credential,
  long-plural  = verifiable credentials
}

\DeclareAcronym{sgx}{
  short = SGX,
  long  = Software Guard Extensions
}

\DeclareAcronym{poa}{
  short = PoA,
  long  = proof of authority
}

\DeclareAcronym{sha256}{
  short = SHA-256,
  long  = Secure Hash Algorithm 256
}

\DeclareAcronym{bft}{
  short = BFT,
  long  = Byzantine fault tolerance
}

\DeclareAcronym{pos}{
  short = BFT,
  long  = proof of stake
}

\begin{document}

\author{
\IEEEauthorblockN{
\begin{tabular*}{\textwidth}{@{\extracolsep{\fill}} >{\centering\arraybackslash}m{0.30\textwidth} >{\centering\arraybackslash}m{0.30\textwidth} >{\centering\arraybackslash}m{0.30\textwidth}}
Ali Irzam Kathia &
Yimika Erinle &
Abylay Satybaldy \\[0.3em]
Paolo Tasca &
Nikhil Vadgama &
Marco Alberto Javarone
\end{tabular*}
}
\\
\IEEEauthorblockA{
Exponential Science Foundation, Lugano, Switzerland
}
}
\title{SoK: A Systematic Bidirectional Literature
Review of AI \texorpdfstring{\&}{\&} DLT Convergence}
\date{}

\maketitle

\begin{abstract}
The integration of Artificial Intelligence (AI) with Distributed Ledger Technology (DLT) has become a growing research area, yet contributions tend to cluster around specific application domains or examine only one direction of the integration, leaving the broader architectural interplay between the two technologies poorly understood. 
This work addresses that gap through a structured, bidirectional review of peer-reviewed studies published between 2020 and 2025. We classify contributions along two directions: AI-enhanced DLT, and DLT-enhanced AI. In the first case, we examine how AI techniques improve DLT systems across five layers: data, network, consensus, execution, and application layers. 
In the second case, we analyse how DLT supports AI systems across five layers: infrastructure, data, model, inference, and application layers, with particular attention to federated learning, model evaluation, and multi-agent coordination. 
The analysis reveals that most works concentrate on a small subset of layers: execution and consensus for AI-enhanced DLT, data and model for DLT-enhanced AI. Other layers remain comparatively neglected. Despite reported improvements in controlled settings, no study demonstrates deployment at production scale, and the field has not yet offered satisfying answers to fundamental questions around scalability, interoperability, and verifiable execution. We argue that progress will require cross-layer co-design and empirical validation in real-world settings.
\end{abstract}
\vspace{1em}

\begin{IEEEkeywords}
Artificial Intelligence, DLT, Taxonomy
\end{IEEEkeywords}

\section{Introduction}
\label{sec:intro}
The integration of \ac{ai} and DLT\footnote{With DLT here we encompass for simplicity all the possible blockchain architectural configurations
and, for the sake of simplicity, also the larger family of distributed ledger technologies, i.e., community
consensus-based distributed ledgers where the storage of data is not based on chains of blocks.} has attracted increasing research attention in recent years. \acs{ai} systems enable data-driven learning and automated decision-making, yet they frequently operate as opaque black boxes, raising concerns regarding transparency, accountability, data provenance, and centralised control. In contrast, DLT networks provide transparent, immutable, and decentralised infrastructures, but face limitations in scalability, computational efficiency, and adaptability. This contrast has motivated research exploring how each technology can address limitations of the other.
Recent studies explore the potential benefits of this integration. On the research front, mechanisms such as proof-of-learning consensus \cite{bravo2019proof}, zero-knowledge machine learning for verifiable inference \cite{Li2024ZKMLSurvey,jagannath2025ai}, and DLT-secured federated learning \cite{Qu2022BlockchainEnabledSurvey,Issa2023BCFLIoT,Zhu2023Blockchain-empoweredDirections} suggest that such combinations may improve aspects of trust, privacy, and coordination in controlled settings. Concurrently, decentralised \acs{ai} agents are increasingly deployed in on-chain environments \cite{Karim2025AIAgents,Raheman2021CryptoAgent}. Industry initiatives such as Bittensor \cite{Bittensor2024}, Fetch.ai \cite{fetch_ai}, and SingularityNET \cite{SingularityNET2018} illustrate the early attempts at integrating token incentives, distributed training, and agent coordination in practice.
Despite this momentum, academic understanding remains fragmented. Research on \acs{ai}-enhanced DLT and DLT-enhanced \acs{ai} often evolves in parallel streams, yielding valuable but siloed contributions. This separation obscures the fundamentally bidirectional nature of the integration, limits comparison across architectural layers and application domains, and limits understanding of how proposed mechanisms scale beyond experimental prototypes. A consolidated and structured overview is therefore necessary to clarify the state of the field and identify shared design patterns and persistent challenges.
This paper addresses these gaps through a structured synthesis of existing research. We review and classify peer-reviewed research along two enhancement directions: \acs{ai} applied to DLT, and DLT applied to \acs{ai}. Within each direction, we analyse on which architectural layer the community focuses the most, the design mechanisms employed, and the maturity of implementation relative to real-world practice. By examining these contributions jointly, we provide a cross-domain synthesis that identifies recurring patterns and open research questions.
In summary, our main contributions are as follows:
\begin{itemize}
    \item \textbf{Bidirectional Synthesis:} We organise research on \acs{ai}-enhanced DLT and DLT-enhanced \acs{ai}, providing a unified perspective across both enhancement directions.  
    \item \textbf{Layer- and Application-Oriented Classification:} We categorise contributions according to architectural layers and application domains, enabling structured comparison across use cases.  
    \item \textbf{Cross-Domain Insights:} We identify cross-domain insights by analysing both directions jointly, identifying common design patterns, complementarities, and divergences.  
    \item \textbf{Gap Analysis:} We identify areas where research remains fragmented or underexplored, highlighting gaps and challenges.   
\end{itemize}

\section{Related Work}
\label{sec:related}

\begin{table*}[!t]
  \centering
  \renewcommand{\arraystretch}{1.2}
    \caption{Overview of related work, indicating for each paper its study type (survey or tool), integration focus (AI-enhanced DLT or DLT-enhanced AI), evaluation approach, covered technical topics, and use cases (\CIRCLE: included, \Circle: not included).}
  \begin{tabular}{r cc cc cc cccc cccccc}
    \toprule
    & \multicolumn{2}{c}{\textbf{Study Type}}
    & \multicolumn{2}{c}{\textbf{Focus}}
    & \multicolumn{2}{c}{\textbf{Evaluation}}
    & \multicolumn{5}{c}{\textbf{Topics}}
    & \multicolumn{5}{c}{\textbf{Use Cases}} \\ 
    \cmidrule(lr){2-3}\cmidrule(lr){4-5}\cmidrule(lr){6-7}
    \cmidrule(lr){8-12}\cmidrule(lr){13-17}
    \textbf{Reference}
    & \rotatebox{90}{Survey}
    & \rotatebox{90}{Tool}
    & \rotatebox{90}{AI‑enhanced DLT}
    & \rotatebox{90}{DLT‑enhanced AI}
    & \rotatebox{90}{Quantitative}
    & \rotatebox{90}{Qualitative}
    & \rotatebox{90}{Security}
    & \rotatebox{90}{Privacy}
    & \rotatebox{90}{Decentralisation}
    & \rotatebox{90}{Accuracy}
    & \rotatebox{90}{Latency}
    & \rotatebox{90}{\acs{iot}}
    & \rotatebox{90}{Transport}
    & \rotatebox{90}{Health}
    & \rotatebox{90}{Finance}
    & \rotatebox{90}{Generic} \\
    \midrule
    \cite{Singh2020BlockIoTIntelligence:Intelligence}
    & \Circle & \CIRCLE
    & \CIRCLE & \CIRCLE
    & \CIRCLE & \CIRCLE
    & \CIRCLE & \CIRCLE & \CIRCLE
    & \CIRCLE & \CIRCLE
    & \CIRCLE & \Circle & \Circle & \Circle & \Circle \\
    \cite{Qu2022Blockchain-enabledSurvey}
    & \CIRCLE & \Circle
    & \Circle & \CIRCLE
    & \Circle & \CIRCLE
    & \CIRCLE & \CIRCLE & \CIRCLE
    & \CIRCLE & \CIRCLE
    & \Circle & \CIRCLE & \CIRCLE & \Circle & \CIRCLE \\
    \cite{Zhu2023Blockchain-empoweredDirections}
    & \CIRCLE & \Circle
    & \Circle & \CIRCLE
    & \Circle & \CIRCLE
    & \CIRCLE & \CIRCLE & \CIRCLE
    & \CIRCLE & \Circle
    & \CIRCLE & \CIRCLE & \CIRCLE & \Circle & \CIRCLE \\
    \cite{Haddad2022SystematicSystems}
    & \CIRCLE & \Circle
    & \Circle & \CIRCLE
    & \Circle & \CIRCLE
    & \CIRCLE & \CIRCLE & \CIRCLE
    & \CIRCLE & \Circle
    & \Circle & \Circle & \CIRCLE & \Circle & \Circle \\
    \cite{Salah2019BlockchainChallenges}
    & \CIRCLE & \Circle
    & \Circle & \CIRCLE
    & \Circle & \CIRCLE
    & \CIRCLE & \CIRCLE & \CIRCLE
    & \Circle & \Circle
    & \Circle & \Circle & \CIRCLE & \Circle & \CIRCLE \\
    \cite{Vilas-Boas2023ConvergenceOpportunities}
    & \CIRCLE & \Circle
    & \Circle & \CIRCLE
    & \Circle & \CIRCLE
    & \CIRCLE & \CIRCLE & \CIRCLE
    & \Circle & \Circle
    & \CIRCLE & \CIRCLE & \Circle & \Circle & \Circle \\
    \cite{Gupta2023InfluencesChains}
    & \Circle & \Circle
    & \CIRCLE & \Circle
    & \CIRCLE & \CIRCLE
    & \CIRCLE & \CIRCLE & \CIRCLE
    & \Circle & \Circle
    & \Circle & \Circle & \Circle & \CIRCLE & \CIRCLE \\
    \cite{Huynh-The2023ArtificialSurvey}
    & \CIRCLE & \Circle      
    & \CIRCLE & \Circle      
    & \Circle & \CIRCLE      
    & \CIRCLE & \CIRCLE & \CIRCLE  
    & \Circle & \Circle
    & \Circle & \Circle & \CIRCLE & \CIRCLE & \CIRCLE \\
    \cite{Ressi2024AI-enhancedOpportunities}
    & \CIRCLE & \Circle
    & \CIRCLE & \Circle
    & \Circle & \CIRCLE
    & \CIRCLE & \CIRCLE & \CIRCLE
    & \Circle & \Circle
    & \CIRCLE & \CIRCLE & \CIRCLE & \CIRCLE & \CIRCLE \\
    \cite{bravo2019proof}
    & \Circle & \CIRCLE
    & \CIRCLE & \Circle
    & \Circle & \CIRCLE
    & \CIRCLE & \Circle & \CIRCLE
    & \Circle & \Circle
    & \Circle & \Circle & \Circle & \Circle & \CIRCLE \\
    \cite{chenli2021dlchain}
    & \Circle & \CIRCLE
    & \CIRCLE & \Circle
    & \CIRCLE & \CIRCLE
    & \CIRCLE & \Circle & \CIRCLE
    & \CIRCLE & \Circle
    & \Circle & \Circle & \Circle & \Circle & \CIRCLE \\
    \cite{zhang2020skychain}
    & \Circle & \CIRCLE
    & \CIRCLE & \Circle
    & \CIRCLE & \CIRCLE
    & \CIRCLE & \Circle & \CIRCLE
    & \Circle & \CIRCLE
    & \Circle & \Circle & \Circle & \Circle & \CIRCLE \\
    \cite{li2023auto}
    & \Circle & \CIRCLE
    & \CIRCLE & \Circle
    & \CIRCLE & \CIRCLE
    & \Circle & \Circle & \CIRCLE
    & \Circle & \CIRCLE
    & \Circle & \Circle & \Circle & \Circle & \CIRCLE \\
    \cite{liu2022permissioned}
    & \CIRCLE & \Circle
    & \CIRCLE & \Circle
    & \CIRCLE & \CIRCLE
    & \CIRCLE & \Circle & \CIRCLE
    & \Circle & \CIRCLE
    & \CIRCLE & \Circle & \Circle & \Circle & \Circle \\
    \cite{salimitari2019ai}
    & \Circle & \CIRCLE
    & \CIRCLE & \Circle
    & \CIRCLE & \CIRCLE
    & \CIRCLE & \Circle & \CIRCLE
    & \CIRCLE & \Circle
    & \CIRCLE & \Circle & \Circle & \Circle & \Circle \\
    \cite{mohammadi2022forkpred}
    & \Circle & \CIRCLE
    & \CIRCLE & \Circle
    & \CIRCLE & \CIRCLE
    & \Circle & \Circle & \CIRCLE
    & \CIRCLE & \Circle
    & \Circle & \Circle & \Circle & \Circle & \CIRCLE \\
    \cite{jagannath2025ai}
    & \Circle & \CIRCLE
    & \Circle & \CIRCLE
    & \CIRCLE & \CIRCLE
    & \CIRCLE & \CIRCLE & \CIRCLE
    & \CIRCLE & \Circle
    & \Circle & \Circle & \Circle & \Circle & \CIRCLE \\
    \cite{scicchitano2020deep}
    & \Circle & \CIRCLE
    & \CIRCLE & \Circle
    & \CIRCLE & \CIRCLE
    & \CIRCLE & \Circle & \CIRCLE
    & \CIRCLE & \Circle
    & \Circle & \Circle & \Circle & \Circle & \CIRCLE \\
    \cite{zhang2023rlamm}
    & \Circle & \CIRCLE
    & \CIRCLE & \Circle
    & \CIRCLE & \CIRCLE
    & \Circle & \Circle & \CIRCLE
    & \Circle & \Circle
    & \Circle & \Circle & \Circle & \CIRCLE & \Circle \\
    \cite{ning2024blockchain}
    & \CIRCLE & \Circle
    & \Circle & \CIRCLE
    & \Circle & \CIRCLE
    & \CIRCLE & \CIRCLE & \CIRCLE
    & \Circle & \Circle
    & \Circle & \Circle & \Circle & \Circle & \CIRCLE \\
    \cite{wu2023survey}
    & \CIRCLE & \Circle
    & \Circle & \CIRCLE
    & \Circle & \CIRCLE
    & \CIRCLE & \CIRCLE & \CIRCLE
    & \Circle & \Circle
    & \CIRCLE & \CIRCLE & \CIRCLE & \Circle & \CIRCLE \\
    \cite{kayikci2024blockchain}
    & \CIRCLE & \Circle
    & \Circle & \CIRCLE
    & \Circle & \CIRCLE
    & \CIRCLE & \CIRCLE & \CIRCLE
    & \Circle & \Circle
    & \Circle & \Circle & \CIRCLE & \CIRCLE & \CIRCLE \\
    \cite{zhu2024survey}
    & \CIRCLE & \Circle
    & \Circle & \Circle
    & \Circle & \CIRCLE
    & \CIRCLE & \CIRCLE & \CIRCLE
    & \Circle & \Circle
    & \Circle & \Circle & \Circle & \Circle & \CIRCLE \\
    \cite{wang2025your}
    & \Circle & \CIRCLE
    & \Circle & \CIRCLE
    & \CIRCLE & \CIRCLE
    & \CIRCLE & \CIRCLE & \CIRCLE
    & \Circle & \Circle
    & \Circle & \Circle & \Circle & \Circle & \CIRCLE \\
    \cite{al2025toward}
    & \CIRCLE & \Circle
    & \Circle & \CIRCLE
    & \Circle & \CIRCLE
    & \CIRCLE & \CIRCLE & \CIRCLE
    & \Circle & \Circle
    & \CIRCLE & \CIRCLE & \CIRCLE & \CIRCLE & \CIRCLE \\
    \cite{Karim2025AIAgents}
    & \CIRCLE & \Circle
    & \CIRCLE & \CIRCLE
    & \Circle & \CIRCLE
    & \CIRCLE & \CIRCLE & \CIRCLE
    & \Circle & \Circle
    & \Circle & \CIRCLE & \Circle & \CIRCLE & \CIRCLE \\
    \bottomrule
  \end{tabular}
  \label{tab:literature-singh}
\end{table*}

\subsection{\acs{ai}-enhanced DLT}

Recent research has increasingly explored the integration of \acs{ai} into DLT systems to enhance their functionality, efficiency, and robustness. By incorporating \acs{ai}-driven methods into DLT systems, these approaches enable more efficient protocol operation, adaptive behaviour under dynamic network conditions, and more effective decision-making across key system layers.

A growing body of research focuses on the use of \acs{ai} to optimise resource management and system performance, particularly in distributed and resource-constrained environments such as the \acf{iiot}. For example, \acs{ai}-driven frameworks have been proposed to improve latency, accuracy, and privacy in DLT-enabled systems, demonstrating advantages over traditional cloud and edge-based architectures~\cite{Singh2020BlockIoTIntelligence:Intelligence}. \acf{rl} has been further applied to dynamically allocate resources and adapt system behaviour under uncertainty, improving reliability and efficiency in heterogeneous network conditions~\cite{liu2022permissioned}. Similarly, \acs{ai}-based anomaly detection mechanisms have been integrated into DLT pipelines to filter abnormal data prior to validation, enhancing system robustness~\cite{salimitari2019ai}.

Within consensus design, \acs{ai} has been explored as a mechanism to redefine how agreement is reached in decentralised systems. Several works propose replacing or augmenting traditional consensus with learning-based paradigms, where computational effort is directed toward solving machine learning tasks. For instance, \enquote{Proof-of-Learning} links consensus participation to model performance, while related approaches leverage deep learning training as a form of useful work embedded within the validation process~\cite{bravo2019proof,chenli2021dlchain}. Complementary approaches employ \acs{rl} to dynamically tune protocol parameters, such as block size and latency, enabling adaptive optimisation of DLT performance in permissioned settings~\cite{li2023auto}.

Beyond protocol optimisation, \acs{ai} techniques have been applied to strengthen DLT security and economic mechanisms. \acf{dl} models have been used to detect anomalous network behaviour and identify potential attacks, improving the resilience of DLT infrastructures \cite{scicchitano2020deep}. In parallel, \acs{ai}-driven approaches have been introduced in \acf{defi}, where \acs{rl}-based \acf{amm} optimise liquidity provision and pricing strategies in \acf{dex}~\cite{zhang2023rlamm}.

At the application layer, \acs{ai}-enhanced DLT architectures have been explored in domains such as supply chain management, where intelligent analytics improve risk assessment, transparency, and operational efficiency~\cite{Gupta2023InfluencesChains}. More broadly, existing surveys highlight the breadth of \acs{ai} integration across consensus, smart-contract security, scalability, and interoperability~\cite{Ressi2024AI-enhancedOpportunities}. However, these analyses remain largely qualitative and lack systematic empirical benchmarking, limiting the comparability and generalisability of reported results.

\subsection{DLT-enhanced \acs{ai}}

The integration of DLT with \acs{ai} systems has been primarily motivated by the need to address challenges related to trust, accountability, and coordination in distributed and multi-party environments. In this context, distributed ledgers facilitate trusted data exchange, coordinated interaction among participants, and verifiable operations through immutable records and programmable logic.

A substantial body of research focuses on the use of DLT to support decentralised model training, particularly in federated learning (FL) settings~\cite{ning2024blockchain,wu2023survey}. Here, the ledger replaces the trusted coordinator. For instance, training rounds are sequenced on-chain, contributions are logged immutably, and reward allocation is governed by smart contract logic rather than by any single participant. This integration enables verifiable logging of updates and transparent reward distribution, thereby mitigating issues such as free-riding and adversarial manipulation. At the same time, incorporating DLT into training pipelines introduces non-trivial system-level trade-offs, including increased latency, higher operational costs, and additional attack surfaces associated with consensus protocols and off-chain dependencies.

Beyond training, DLT has been explored as a mechanism for strengthening the integrity and governance of end-to-end \acs{ai} pipelines~\cite{kayikci2024blockchain,zhu2024survey,wang2025your}. In this context, distributed ledgers are used to record the provenance of datasets, model artefacts, and inference outputs, enabling traceability across the lifecycle of \acs{ai} systems. Such capabilities are particularly relevant in high-stakes domains where auditability and reproducibility are essential. However, practical deployment remains constrained by challenges related to data standardisation, storage efficiency, and the scalability of on-chain recording mechanisms.

More recent research extends this perspective toward fully decentralised \acs{ai} ecosystems, where DLT functions as a coordination backbone for complex, multi-actor environments~\cite{al2025toward,Karim2025AIAgents}. In these settings, smart contracts facilitate interactions among autonomous agents, service providers, and data contributors within shared marketplaces. This enables the emergence of programmable incentive structures and composable services, supporting more flexible and open \acs{ai} architectures. Nevertheless, these systems introduce new challenges, including the need for robust interoperability across platforms, alignment of economic incentives with system-level objectives, and compliance with evolving regulatory frameworks.

\subsection{Addressing Literature Gaps}
\label{sec:gaps-in-literature}

Although prior research demonstrates promising intersections between \acs{ai} and DLT, most studies focus on specific topics, such as security, privacy, or particular application domains, without systematically connecting insights across enhancement directions (\autoref{tab:literature-singh}). Existing surveys typically remain domain-bound or adopt a single-direction perspective, leaving the bidirectional and architectural interplay between \acs{ai} and DLT insufficiently examined. This fragmentation limits comparability across studies and obscures recurring design patterns, trade-offs, and challenges that emerge across application contexts.

This study addresses these limitations by jointly analysing both enhancement directions: \acs{ai} applied to DLT and DLT applied to \acs{ai}. We introduce a unified classification framework that organises contributions along three dimensions: application domain, architectural layer, and design mechanism. This structured approach enables cross-domain comparison, reveals recurring complementarities and tensions, and surfaces persistent gaps in current research. By moving beyond fragmented, domain-specific analyses toward an integrated system-level synthesis, we provide a clearer and more comprehensive understanding of \acs{ai}–DLT convergence, and outline concrete directions for future research and practical deployment.

\section{Generalised AI Layer Stack}
\label{sec:ai_layer}

AI systems can be organised into a five-layer architecture reflecting common abstractions in AI system design. Prior research shows that AI development involves interdependent components, including data management, model construction, computational infrastructure, and deployment operations \cite{sculley2015hidden, amershi2019software}. The five-layer architecture adopted in this study organises these abstractions into a consistent framework and provides a basis for comparison with the layered DLT stack.

\subsection{Infrastructure Layer}

The infrastructure layer provides the computational foundation for AI workloads. It encompasses hardware accelerators \cite{jouppi2017datacenter}, distributed training systems \cite{dean2012large}, as well as storage and networking systems that support large-scale model training and serving. This layer also includes orchestration and runtime systems responsible for resource management and scaling. Infrastructure defines the baseline computational capacity and reliability upon which higher-level AI functions depend.

\subsection{Data Layer}

The data layer governs dataset acquisition, preprocessing, and governance. Robust AI systems require mechanisms to track data lineage, transformations, and versioning to ensure reproducibility and accountability \cite{buneman2001why}. Automated validation, schema enforcement, and constraint checking further enhance consistency in data pipelines \cite{schelter2018automating}. This layer governs how data is structured and prepared for model use.

\subsection{Model Layer}

The model layer encompasses model design, training, optimisation, evaluation, and lifecycle management. It includes hyperparameter tuning, checkpointing, and model selection processes. Beyond purely technical aspects, this layer incorporates governance considerations such as documentation, explainability, ownership, licensing, and compliance, particularly in multi-stakeholder or regulated settings \cite{mitchell2019modelcards}. The model layer defines how models are created, maintained, and controlled within AI systems.

\subsection{Inference Layer}

The inference layer comprises mechanisms for executing and serving trained models in production environments. Key considerations include runtime optimisation, latency, and monitoring. Some approaches enable verifiable and privacy-preserving inference using encrypted computation, secure multiparty computation, or trusted execution environments \cite{gilad2016cryptonets, phuong2019privacy}. This layer ensures that model outputs are delivered efficiently, securely, and, where required, with integrity or confidentiality guarantees.

\subsection{Application Layer}

The application layer integrates AI capabilities into end-to-end operational workflows. It includes orchestration, governance controls, incentive mechanisms, and auditability surrounding model usage \cite{dignum2019responsible}. This layer defines how AI systems are invoked, constrained, supervised, and evaluated in real-world contexts, serving as the interface between AI systems and real-world use cases.

This five-layer architecture provides a structural representation of AI systems across development and deployment stages. Its conceptual alignment with DLT’s layered architecture enables cross-layer comparison and supports mapping of enhancement mechanisms in both directions of AI–DLT integration.

\section{Generalised DLT Layer Stack}
\label{sec:blockchain_layer}
DLT technology can be represented as a layered system that abstracts the core functions of distributed ledgers into a structured architecture. Prior studies propose different perspectives on this abstraction, yet the literature converges on a common set of foundational layers \cite{birje2022review, tabatabaei2023understanding}. By synthesising these insights, we propose a five-layer DLT architecture comprising data, network, consensus, execution, and application, providing a unified basis for comparison with parallel AI architectures.

\subsection{Data Layer}

The data layer defines how DLT state and transaction history are represented, stored, and made available to participants. It encompasses the structure of blocks, transactions, and world state, as well as the mechanisms used to ensure integrity and retrievability, such as Merkle-based commitments and indexing structures. This layer also governs how data is accessed by full nodes, light clients, and external systems.

Data availability is a central concern at this layer, particularly in modular and rollup-based architectures, where nodes must verify that required data is published and accessible without necessarily storing it in full~\cite{saif2024survey}. Design choices at this level determine how efficiently the state can be retrieved, verified, and maintained over time. The data layer, therefore, establishes the foundation upon which execution operates and consensus validates system state.

\subsection{Network Layer}

The network layer enables communication between distributed nodes and supports the propagation of transactions and blocks across the system. It encompasses peer discovery, message dissemination, and synchronisation mechanisms that allow nodes to maintain a consistent view of the DLT \cite{birje2022review}.

This layer determines how efficiently information flows through the system, influencing latency, throughput, and resilience to network-level attacks. Gossip-based protocols are commonly used to balance scalability and robustness, though they introduce trade-offs in bandwidth usage and propagation delay \cite{xu2017taxonomy}. The network layer thus provides the communication substrate that allows consensus and execution to operate over a distributed set of participants.
\subsection{Consensus Layer}
The consensus layer establishes agreement among distributed nodes on the validity and ordering of transactions. It defines the rules by which blocks are proposed, validated, and finalised, ensuring that all participants share a consistent view of the ledger state. Beyond safety and liveness, consensus protocols also determine key system properties such as throughput, latency, and fault tolerance \cite{bano2019sok,xu2023survey}.

Classical protocols such as \acf{pbft} inspired many permissioned systems, while Nakamoto consensus introduced probabilistic finality through proof-of-work. Subsequent designs, proof-of-stake, committee-based BFT, and hybrid protocols, aim to improve energy efficiency, scalability, and security against adversarial behaviour~\cite{xu2023survey}. Recent work explores sharding, \acf{dag}-based consensus, and \acf{vrf} as mechanisms to further enhance performance and decentralisation~\cite{bano2019sok}.

\subsection{Execution Layer}

The execution layer defines how DLT transactions are interpreted and deterministically evaluated to produce state transitions. By ensuring that all nodes compute identical results from the same ordered inputs and prior state, this layer operationalises the DLT as a replicated state machine \cite{alzhrani2022taxonomy, bellaj2024drawing}.

At this layer, smart contracts and transaction logic specify the rules governing assets, identities, and interactions, enabling conditional and automated system behaviour. The execution layer, therefore, translates consensus-confirmed transactions into state updates and provides the computational foundation for decentralised applications. 

\subsection{Application Layer}
The application layer represents the user-facing functionality of DLT systems. It includes smart contracts, decentralised applications (dApps), and the services that leverage the underlying consensus and data layers to deliver functionality such as decentralised finance, digital identity, supply chain management, and governance~\cite{alzhrani2022taxonomy}.

Application-layer issues, such as usability, composability, and interoperability, strongly influence adoption, while the security of smart contracts remains a central concern. By building directly on the guarantees of consensus and data integrity, the application layer enables diverse ecosystems of decentralised services and provides the most direct avenue for integrating DLT with \acs{ai} systems.

\section{Methodology}
\label{sec:method}

\subsection{Objectives and Scope}

The objective of this study is to provide a structured classification of research at the intersection of \acs{ai} and DLT. Instead of focusing on individual application domains, we organise contributions according to the layered reference models introduced in \autoref{sec:ai_layer} and \autoref{sec:blockchain_layer}. This taxonomy~\cite{tascatax} enables comparison of how \acs{ai} enhances DLT and how DLT enhances \acs{ai} across their respective functional stacks.

The review is restricted to some peer-reviewed research published between 2020 and 2025 that explicitly addresses \acs{ai}--DLT integration. We include studies that propose concrete mechanisms, architectural frameworks, or implemented systems. Conceptual or speculative studies lacking technical or empirical grounding were excluded.

\subsection{Sources and Coverage}

Relevant studies were identified through structured searches in Google Scholar, IEEE Xplore, and the ACM Digital Library. To enhance coverage and reduce omission bias, the initial corpus was expanded using backward reference tracing and forward citation analysis. The final dataset is limited to peer-reviewed literature to ensure methodological quality and comparability. The final dataset comprises $53$ studies after applying the inclusion and exclusion criteria.

\subsection{Inclusion and Exclusion Criteria}
\label{sec:inclusion-criteria}

To ensure relevance and comparability, explicit inclusion and exclusion criteria were applied during screening.

\subsubsection*{Inclusion criteria}
\begin{itemize}
    \item Peer-reviewed journal articles, systematic reviews, surveys, or full-length conference papers presenting technical contributions.
    \item Explicit examination of \acs{ai}--DLT integration in either the \acs{ai}-enhanced DLT or DLT-enhanced \acs{ai} direction.
    \item Specification of enhancement mechanisms, architectures, or experimentally evaluated systems.
    \item Publication date between 2020 and 2025.
\end{itemize}

\subsubsection*{Exclusion criteria}
\begin{itemize}
    \item Studies addressing \acs{ai} or DLT independently without integrative enhancement.
    \item Duplicate publications or extended versions of already included works.
    \item High-level vision papers, position statements, or conceptual discussions lacking methodological, technical, or empirical substantiation.
\end{itemize}

\subsection{Data Extraction and Classification}
\label{sec:data-extraction}

For each included study, relevant metadata were extracted using a structured classification template. The following dimensions were recorded:

\begin{itemize}
    \item \textbf{Bibliographic information:} Authors, publication year, and venue.
    \item \textbf{Enhancement direction:} \acs{ai} enhancing DLT or DLT enhancing \acs{ai}.
    \item \textbf{Enhancement layer:} The specific layer of the \acs{ai} stack (\autoref{sec:ai_layer}) or DLT stack (\autoref{sec:blockchain_layer}) targeted by the contribution.
    \item \textbf{Domain:} The application domain addressed (e.g., consensus optimisation, \acs{dex} design, \acs{fl}, provenance management).
    \item \textbf{Mechanism:} The main mechanism (e.g., \acs{rl}, \acs{zkp}, smart contracts, token incentives).
    \item \textbf{Application category:} The objective or impact domain (e.g., security, privacy, scalability, governance, interoperability, auditability).
    \item \textbf{Deployment maturity:} The implementation level, categorised as prototype (experimentally validated in simulation or testbed) or deployed (operational in real-world settings).
\end{itemize}

This structured classification maintains consistency across both directions and enables cross-layer comparison. The resulting taxonomy is used for the comparative insights and gap analysis presented in \autoref{sec:discussion}.

\section{Results}
\label{sec:results}

\subsection{AI-enhanced DLT}

\begin{table*}[htbp]
\centering
\tiny
\setlength{\tabcolsep}{2.95pt}
\renewcommand{\arraystretch}{1.35}
\caption{Classification of AI-enhanced DLT studies by enhancement layer, DLT domain, AI mechanism, and application category. The table highlights how AI methods are applied to improve specific components of the DLT stack and their associated application areas.}
\resizebox{\textwidth}{!}{%
\begin{tabular}{@{}lll*{7}{c}c@{}}
\toprule
\textbf{\makecell[l]{DLT\\Layer}} & \textbf{\makecell[l]{DLT\\ Domain}} & \textbf{AI Mechanism} &
\rotatebox{90}{\textbf{Security}} &
\rotatebox{90}{\textbf{Scalab.}} &
\rotatebox{90}{\textbf{Interop.}} &
\rotatebox{90}{\textbf{Auditab.}} &
\rotatebox{90}{\textbf{Sustain.}} &
\rotatebox{90}{\textbf{Incentiv.}} &
\rotatebox{90}{\textbf{Explain.}} &
\textbf{Study} \\
\midrule
\multirow{1}{*}{\textbf{DATA}}
 & Data Storage   & ML Model-based Indexing  &
 \smallemptycirc& \smallfullcirc & \smallemptycirc &
 \smallemptycirc & \smallemptycirc & \smallemptycirc & \smallemptycirc &
 \cite{zhang2024cole} \\[-1pt]
\midrule
\multirow{4}{*}{\textbf{NET}}
 & Network Optimisation & ML Models (MLR, KMeans, RF) &
 \smallfullcirc & \smallemptycirc & \smallemptycirc &
 \smallemptycirc & \smallemptycirc & \smallemptycirc & \smallemptycirc &
 \cite{maaroufi2021bcool} \\[-1pt]
 & Anomaly \& Fraud Detection & Semi-supervised AutoEncoder &
 \smallemptycirc & \smallfullcirc & \smallemptycirc &
 \smallemptycirc & \smallemptycirc & \smallemptycirc & \smallemptycirc &
 \cite{kim2022machine} \\[-1pt]
& Anomaly \& Fraud Detection & CNN–LSTM Models  &
 \smallfullcirc & \smallemptycirc & \smallemptycirc &
 \smallemptycirc & \smallemptycirc & \smallemptycirc & \smallemptycirc &
 \cite{abdulrazzaq2020decentralized} \\[-1pt]
& Network Congestion Control  & DDPG Algorithm   &
 \smallemptycirc & \smallfullcirc & \smallemptycirc &
 \smallemptycirc & \smallemptycirc & \smallemptycirc & \smallemptycirc &
 \cite{ning2021intelligent} \\[-1pt]
\midrule
\multirow{7}{*}{\textbf{CON}}
 & Consensus Mechanism & RL &
   \smallemptycirc & \smallfullcirc & \smallemptycirc &
   \smallemptycirc & \smallemptycirc & \smallemptycirc & \smallemptycirc &
   \cite{villegas2025adaptive} \\[-1pt]
 & Consensus Mechanism & Deep Reinforcement Learning (DRL) &
   \smallfullcirc & \smallfullcirc & \smallemptycirc &
   \smallemptycirc & \smallemptycirc & \smallemptycirc & \smallemptycirc &
   \cite{goh2022secure} \\[-1pt]
 & Consensus Mechanism & DL PoUW (DLchain) &
   \smallfullcirc & \smallemptycirc & \smallemptycirc &
   \smallemptycirc & \smallfullcirc & \smallemptycirc & \smallemptycirc &
   \cite{chenli2021dlchain} \\[-1pt]
 & Consensus Mechanism & ML Models (DL, RL, GANs) &
   \smallfullcirc & \smallemptycirc & \smallemptycirc &
   \smallemptycirc & \smallemptycirc & \smallemptycirc & \smallemptycirc &
   \cite{venkatesan2024blockchain} \\[-1pt]
 & Consensus Mechanism & Multi-agent Deep RL &
   \smallemptycirc & \smallfullcirc & \smallemptycirc &
   \smallemptycirc & \smallemptycirc & \smallemptycirc & \smallemptycirc &
   \cite{li2023auto} \\[-1pt]
 & Consensus Mechanism & PoLe &
   \smallemptycirc & \smallemptycirc & \smallemptycirc &
   \smallemptycirc & \smallfullcirc & \smallemptycirc & \smallemptycirc &
   \cite{lan2021proof} \\[-1pt]
 & Sharding Protocol & DRL Dynamic Sharding &
   \smallemptycirc & \smallfullcirc & \smallemptycirc &
   \smallemptycirc & \smallemptycirc & \smallemptycirc & \smallemptycirc &
   \cite{zhang2020skychain} \\[-1pt]
 & Fork Monitoring     & ML Fork Prediction &
   \smallfullcirc & \smallemptycirc & \smallemptycirc &
   \smallemptycirc & \smallemptycirc & \smallemptycirc & \smallemptycirc &
   \cite{mohammadi2022forkpred} \\[-1pt]
\midrule
\multirow{8}{*}{\textbf{EXE}}
 & Smart-Contract  & LLMs &
   \smallfullcirc & \smallemptycirc & \smallemptycirc &
   \smallemptycirc & \smallemptycirc & \smallemptycirc & \smallemptycirc &
   \cite{chen2025cryptic} \\[-1pt]
 & Smart-Contract & DL Models   &
   \smallfullcirc & \smallemptycirc & \smallemptycirc &
   \smallemptycirc & \smallemptycirc & \smallemptycirc & \smallemptycirc &
   \cite{tang2023deep} \\[-1pt]
 & Smart-Contract & ML and DL Models &
   \smallfullcirc & \smallemptycirc & \smallemptycirc &
   \smallemptycirc & \smallemptycirc & \smallemptycirc & \smallemptycirc &
   \cite{shah2023deep} \\[-1pt]
 & Smart-Contract & Multimodal DL Framework &
   \smallfullcirc & \smallemptycirc & \smallemptycirc &
   \smallemptycirc & \smallemptycirc & \smallemptycirc & \smallemptycirc & 
   \cite{deng2023smart} \\[-1pt]
 & Smart-Contract & LLMs &
   \smallfullcirc & \smallemptycirc & \smallemptycirc &
   \smallemptycirc & \smallemptycirc & \smallemptycirc & \smallemptycirc &
   \cite{khalid2025evaluating} \\[-1pt]
  & Smart-Contract & NLP Analysis &
   \smallfullcirc & \smallemptycirc & \smallemptycirc &
   \smallemptycirc & \smallemptycirc & \smallemptycirc & \smallemptycirc &
   \cite{raja2020ai} \\[-1pt]
  & Smart-Contract & Reinforced LLM Framework &
   \smallfullcirc & \smallemptycirc & \smallemptycirc &
   \smallemptycirc & \smallemptycirc & \smallemptycirc & \smallfullcirc  &
   \cite{yu2025smart} \\[-1pt]
  & Smart-Contract & GNN &
   \smallfullcirc & \smallemptycirc & \smallemptycirc &
   \smallemptycirc & \smallemptycirc & \smallemptycirc & \smallemptycirc &
   \cite{liu2023smart} \\[-1pt]
  & Smart-Contract & LLMs &
   \smallemptycirc & \smallemptycirc & \smallemptycirc &
   \smallfullcirc  & \smallemptycirc & \smallemptycirc & \smallemptycirc &
   \cite{chen2025suigpt} \\[-1pt]
\midrule
\multirow{4}{*}{\textbf{APP}}
  & Decentralised Identity & NLP Analysis &
   \smallfullcirc & \smallemptycirc & \smallemptycirc &
   \smallemptycirc & \smallemptycirc & \smallemptycirc & \smallemptycirc &
   \cite{chinnasamy2025blockchain} \\[-1pt]
 & Decentralised Exchange & RL-Steered AMM &
   \smallemptycirc & \smallemptycirc& \smallemptycirc &
   \smallemptycirc & \smallemptycirc & \smallfullcirc & \smallemptycirc &
   \cite{churiwala2023qlammp}\\[-1pt]
 & Oracles \& Data Feeds & LLM-based Oracle Framework &
   \smallemptycirc & \smallemptycirc& \smallfullcirc &
   \smallemptycirc & \smallemptycirc & \smallemptycirc & \smallemptycirc &
   \cite{zeng2025connecting}\\[-1pt]
& Decentralised Finance & DL Models, NLP, XAI Methods &
   \smallfullcirc & \smallemptycirc & \smallemptycirc &
   \smallemptycirc & \smallemptycirc & \smallemptycirc & \smallfullcirc &
   \cite{idowuleveraging}\\[-1pt]
\bottomrule
\end{tabular}
}
\label{tab:ai_enhanced_blockchain}
\end{table*}

\subsubsection{Scope, Trends, and Methods}

\begin{table*}[h]
\centering
\small
\setlength{\tabcolsep}{5pt}
\renewcommand{\arraystretch}{1.05}
\caption{Distribution of AI-enhanced DLT studies across DLT layers.}
\begin{tabular}{@{} l c c p{8cm} @{}}
\toprule
\makecell{\textbf{Enhanced}\\\textbf{DLT Layer}} &
\makecell{\textbf{No. of}\\\textbf{Studies}} &
\textbf{(\%)} &
\textbf{Application Domain} \\
\midrule
Execution   & 9 & \cellcolor{g4}{34.6} & Smart contract vulnerability detection, bytecode auditing \\
Consensus   & 8 & \cellcolor{g3}{30.8} & Throughput optimization, trust, sustainability \\
Network     & 4 & \cellcolor{o1}{15.4} & Congestion control, anomaly detection, resource allocation \\
Application & 4 & \cellcolor{o1}{15.4} & Oracle intelligence, DID, DEX/AMM, DeFi analytics \\
Data        & 1 & \cellcolor{r3}{3.8}  & Learned data storage, scalability \\
\bottomrule
\end{tabular}
\label{tab:ai_enhanced_blockchain_dist}
\end{table*}

A total of 26 studies were identified where AI techniques were implemented in prototype or deployed settings to enhance DLT systems. Each study was classified according to the enhanced DLT layer, the corresponding AI layer involved, and the primary mechanism applied. \autoref{tab:ai_enhanced_blockchain} summarises the classification results.

As shown in \autoref{tab:ai_enhanced_blockchain_dist}, most contributions targeted the Execution layer (9; 34.6\%), and the Consensus layer (8; 30.8\%). Fewer studies addressed the Network (4; 15.4\%), Application (4; 15.4\%), and Data (1; 3.8\%) layers. 

Across layers, \acs{ai} was predominantly applied at the model layer, with a smaller subset leveraging inference-layer systems such as \acf{llm}s \cite{zeng2025connecting, chen2025cryptic}. The most frequently employed \acs{ai} techniques were \acs{rl}, \acs{dl}, and \acs{llm}-based models. \acs{rl} was primarily used for consensus optimisation and system tuning \cite{goh2022secure, li2023auto, villegas2025adaptive}. \acs{dl} architectures supported anomaly detection, vulnerability analysis, and traffic classification \cite{tang2023deep, deng2023smart, liu2023smart}. \acs{llm}s were applied to smart-contract analysis, vulnerability detection, and explainability \cite{chen2025cryptic, yu2025smart, zeng2025connecting}.

In terms of application domain, 65.4\% of studies addressed security-related objectives, followed by 26.9\% focusing on scalability. Smaller proportions examined explainability and sustainability (7.7\% each), while incentives, auditability, and interoperability were each covered by 3.8\% of the reviewed works. These categories are multi-label and therefore not mutually exclusive.

\subsubsection{Mechanisms and Implementation Focus}

Beyond aggregate trends, the identified mechanisms show how AI techniques interact with DLT components.

\begin{itemize}

\item \textbf{\acs{rl} for control and optimisation:}  
Deep and multi-agent \acs{rl} methods were employed to auto-tune consensus and system parameters in permissioned networks~\cite{li2023auto, goh2022secure, villegas2025adaptive}, optimise sharding~\cite{zhang2020skychain} and resource allocation in mobile DLT settings~\cite{ning2021intelligent}, and adapt \acs{amm} parameters such as trading fees~\cite{churiwala2023qlammp}.

\item \textbf{\acs{dl} for detection and analytics:}  
\acf{cnn} and \acf{lstm} architectures, together with autoencoder-based models, were applied to detect \acf{p2p} traffic anomalies~\cite{kim2022machine}. Transformer, multimodal, and \acf{gnn} architectures were used to identify smart-contract vulnerabilities and malicious patterns~\cite{tang2023deep, deng2023smart, liu2023smart}.

\item \textbf{LLMs and generative tooling:}  
LLMs supported vulnerability detection and explainability~\cite{chen2025cryptic, yu2025smart}, assisted contract development, and were also used in oracle pipelines with commit–reveal aggregation~\cite{zeng2025connecting}.

\item \textbf{\acs{ai}-based consensus mechanisms:}  
\acf{powu}~\cite{chenli2021dlchain} and \acf{pole}~\cite{lan2021proof} embed deep learning training directly into consensus, where neural network computation serves as proof-of-work and model verification ensures result integrity.

\item \textbf{Classical \acs{ml} and hybrid approaches:}  
Supervised and ensemble ML methods addressed network congestion control and role-based access management~\cite{maaroufi2021bcool, chinnasamy2025blockchain}. Hybrid approaches combined learned models with rule-based validators to enhance robustness and precision~\cite{liu2023smart}. Learned data structures further improved on-chain efficiency, where model-based indexing improved storage and query efficiency~\cite{zhang2024cole}.

\end{itemize}

Evaluation methodologies primarily relied on simulation, emulation, or controlled testbeds built on platforms. One study analysed Bitcoin mainnet traffic for anomaly detection~\cite{kim2022machine}. Reported performance metrics included throughput and latency measures, including \acf{tps}, and task-specific performance metrics for optimisation tasks, and security indicators such as precision, recall, and F1-scores for vulnerability detection~\cite{tang2023deep, deng2023smart, liu2023smart}. Oracle aggregation mechanisms commonly employed commit–reveal schemes combined with threshold or committee-based voting prior to on-chain actuation~\cite{zeng2025connecting}. No study reported large-scale production deployment.

\subsection{DLT-enhanced AI}

\begin{table*}[htbp]
\centering
\tiny
\setlength{\tabcolsep}{2.95pt}
\renewcommand{\arraystretch}{1.15}
\caption{Classification of DLT-enhanced AI studies by enhancement layer, AI domain, DLT mechanism, and application category. The table highlights how DLT methods are applied to improve specific components of the AI stack and their associated application areas.}  
\label{tab:blockchain_enhanced_AI}
\resizebox{\textwidth}{!}{%
\begin{tabular}{@{}llll*{6}{c}l@{}}
\toprule
\textbf{AI Layer} &
\textbf{AI Domain} &
\textbf{\makecell[l]{DLT Mechanism}} &
\rotatebox{90}{\textbf{Trust}} &
\rotatebox{90}{\textbf{Privacy}} &
\rotatebox{90}{\textbf{Incent.}} &
\rotatebox{90}{\textbf{Interop.}} &
\rotatebox{90}{\textbf{Explain.}} &
\rotatebox{90}{\textbf{Security}} &
\rotatebox{90}{\textbf{Auditab.}} &
\textbf{Study} \\
\midrule
\multirow{2}{*}{\textbf{INFRA}} 
&
Decentralised Compute   & Public DLT, Staking &
\smallfullcirc & \smallemptycirc & \smallfullcirc & \smallemptycirc & \smallemptycirc & \smallemptycirc & \smallemptycirc &
\cite{wang2025aiarena} \\[-1pt]
&
Model Deployment  & Hybrid DLT &
\smallemptycirc & \smallemptycirc & \smallemptycirc & \smallemptycirc & \smallemptycirc & \smallemptycirc & \smallfullcirc &
\cite{tulkinbekov2025doctrina} \\[-1pt]
\midrule
\multirow{9}{*}{\textbf{DATA}}
& FL Data Sharing & DIDs, Verifiable Credentials &
\smallfullcirc  & \smallfullcirc & \smallemptycirc & \smallemptycirc & \smallemptycirc & \smallemptycirc & \smallemptycirc &
\cite{papadopoulos2021privacy} \\[-1pt]
& FL Data Sharing & Proof of Contribution (PoC) &
\smallemptycirc & \smallfullcirc & \smallemptycirc & \smallemptycirc & \smallemptycirc & \smallfullcirc & \smallemptycirc &
\cite{guo2025verifiable} \\[-1pt]
& FL Data Sharing & Consortium DLT &
\smallemptycirc & \smallfullcirc & \smallemptycirc & \smallemptycirc & \smallemptycirc & \smallemptycirc & \smallemptycirc &
\cite{durga2022fled} \\[-1pt]
&
AI Data Pipeline  & ZKPs, Smart Contract, IPFS &
\smallemptycirc & \smallemptycirc & \smallemptycirc & \smallemptycirc & \smallfullcirc & \smallfullcirc  & \smallemptycirc &
\cite{kumar2023blockchain} \\[-1pt]
& 
AI Data Pipeline & Smart Contract, PoA &
\smallfullcirc & \smallemptycirc & \smallemptycirc & \smallemptycirc & \smallemptycirc & \smallemptycirc & \smallemptycirc &
\cite{shahbazi2021blockchain} \\[-1pt]
&
AI Data Pipeline & ZKPs, Smart Contract  &
\smallemptycirc & \smallemptycirc & \smallemptycirc & \smallemptycirc & \smallemptycirc & \smallfullcirc & \smallemptycirc &
\cite{lin2023blockchain} \\[-1pt] 
& 
Data/Model Provenance & Smart Contract, IPFS &
\smallemptycirc & \smallemptycirc & \smallemptycirc & \smallemptycirc & \smallemptycirc & \smallfullcirc & \smallfullcirc &
\cite{witanto2022toward} \\[-1pt]
& 
Data/Model Provenance & Smart Contract, Permissioned DLT &
\smallemptycirc & \smallemptycirc & \smallemptycirc & \smallemptycirc & \smallemptycirc & \smallfullcirc & \smallfullcirc &
\cite{soldatos2021blockchain} \\[-1pt]
& 
Data/Model Provenance & Smart Contract, IPFS &
\smallemptycirc & \smallemptycirc & \smallemptycirc & \smallemptycirc & \smallemptycirc & \smallfullcirc & \smallfullcirc &
\cite{olawale2024cybersecurity} \\[-1pt]
& 
AIGC Provenance & Smart Contract, NFT &
\smallfullcirc & \smallemptycirc & \smallemptycirc & \smallemptycirc & \smallemptycirc & \smallemptycirc& \smallfullcirc &
\cite{fitriawijaya2024integrating} \\[-1pt]
\midrule
\multirow{6}{*}{\textbf{MODEL}} 
&
Model Training/Evaluation & Smart Contract, IPFS  &
\smallfullcirc & \smallemptycirc & \smallemptycirc & \smallemptycirc & \smallemptycirc & \smallfullcirc & \smallemptycirc &
\cite{jiang2023blockchain} \\ [-1pt]
&
FL Training/Aggregation & PoS-BFT, ZKPs  &
\smallemptycirc & \smallfullcirc & \smallemptycirc & \smallemptycirc & \smallemptycirc & \smallemptycirc & \smallemptycirc &
\cite{kr2025blockchain} \\ [-1pt]
&
FL Training/Aggregation & PoTC  &
\smallfullcirc & \smallemptycirc & \smallemptycirc & \smallemptycirc & \smallemptycirc & \smallfullcirc & \smallemptycirc &
\cite{bi2022iot} \\ [-1pt]
&
FL Training/Aggregation & SHA-256, Smart Contract  &
\smallemptycirc & \smallfullcirc & \smallemptycirc & \smallemptycirc & \smallfullcirc & \smallemptycirc & \smallemptycirc &
\cite{bhardwaj2025explainable} \\  [-1pt]
&
FL Training/Aggregation & Smart Contract  &
\smallfullcirc & \smallemptycirc & \smallemptycirc & \smallemptycirc & \smallemptycirc & \smallemptycirc & \smallfullcirc &
\cite{zerka2020blockchain} \\  [-1pt]
&
Cross-chain FL & Consortium Blokchain (PBFT)  &
\smallemptycirc & \smallemptycirc & \smallemptycirc & \smallfullcirc & \smallemptycirc & \smallemptycirc & \smallemptycirc &
\cite{kang2022communication} \\[-1pt] 
\midrule
\multirow{4}{*}{\textbf{INFER}}
& 
LLM Inference  & Smart Contract&
\smallemptycirc & \smallemptycirc & \smallemptycirc & \smallemptycirc & \smallemptycirc & \smallemptycirc & \smallfullcirc &
\cite{park2024brain} \\[-1pt]
& 
Explainable AI & Smart Contract, Proof of Authority (PoA)  &
\smallemptycirc & \smallemptycirc & \smallemptycirc & \smallemptycirc & \smallfullcirc & \smallemptycirc & \smallfullcirc &
\cite{paulin2023histotrust} \\[-1pt]
& 
Explainable AI & Smart Contract, PoI  &
\smallfullcirc& \smallemptycirc & \smallemptycirc & \smallemptycirc & \smallemptycirc & \smallemptycirc & \smallfullcirc &
\cite{muhammad2025blockchain} \\[-1pt]
& 
Federated Learning & Smart Contract, Permissioned DLT  &
\smallemptycirc & \smallemptycirc & \smallemptycirc & \smallemptycirc & \smallemptycirc  & \smallemptycirc & \smallfullcirc &
\cite{drungilas2021towards} \\[-1pt]
\midrule
\multirow{5}{*}{\textbf{APP}}
&
AI Model Marketplace & Smart Contract  &
\smallemptycirc & \smallemptycirc & \smallfullcirc & \smallemptycirc & \smallemptycirc & \smallemptycirc & \smallemptycirc &
\cite{harris2020analysis} \\ [-1pt]
& 
AI Model Marketplace & Smart Contract, NFT, DAG &
\smallfullcirc & \smallemptycirc & \smallemptycirc & \smallemptycirc & \smallemptycirc & \smallfullcirc & \smallemptycirc &
\cite{battah2022blockchain} \\[-1pt]
& 
AIGC Marketplace  & Smart Contract, IPFS &
\smallfullcirc & \smallemptycirc & \smallemptycirc & \smallemptycirc & \smallemptycirc & \smallemptycirc & \smallemptycirc &
\cite{truong2024trust} \\[-1pt]
& 
Multi-Agent Systems & Proof-of-Thought (PoT) &
\smallemptycirc & \smallemptycirc & \smallemptycirc & \smallemptycirc & \smallemptycirc & \smallfullcirc & \smallemptycirc &
\cite{chen2024blockagents} \\[-1pt]
& 
Multi-Agent Systems & Smart Contract &
\smallemptycirc & \smallemptycirc & \smallemptycirc & \smallemptycirc & \smallemptycirc & \smallfullcirc & \smallemptycirc &
\cite{mhamdi2022trust} \\[-1pt]
\bottomrule
\end{tabular}
}
\end{table*}

\subsubsection{Scope, Trends, and Methods}

A total of 27 publications were identified in which DLT mechanisms were implemented in prototype or deployed settings to enhance AI systems. Each study was classified according to the enhanced AI layer, the corresponding DLT layer involved, and the primary mechanism applied. \autoref{tab:blockchain_enhanced_AI} summarises the classification results.

As shown in \autoref{tab:blockchain_enhanced_AI_dist}, most contributions addressed the Data layer (9; 33.3\%) and the Model layer (7; 25.9\%). The remaining studies focused on the Application (5; 18.5\%), Inference (4; 14.8\%), and Infrastructure (2; 7.4\%) layers.

At the Data and Model layers, DLT was predominantly employed to support federated learning workflows~\cite{guo2025verifiable, kr2025blockchain, zerka2020blockchain}, as well as to secure data provenance, integrity, and pipeline traceability~\cite{witanto2022toward, olawale2024cybersecurity, soldatos2021blockchain}. At the Inference layer, DLT-based logging, attestation, and immutable record-keeping mechanisms supported model explanation, traceability, and auditability~\cite{paulin2023histotrust, muhammad2025blockchain}. Within the AI application layer, decentralised marketplaces enabled model exchange and monetisation through DLT execution mechanisms, particularly smart contracts and \acs{nft}s~\cite{battah2022blockchain, truong2024trust}. Multi-agent systems further leveraged DLT-based trust, reputation, and coordination mechanisms to facilitate secure collaboration among distributed AI agents~\cite{mhamdi2022trust, chen2024blockagents}.

\begin{table*}[h!]
\centering
\small
\setlength{\tabcolsep}{5pt} 
\renewcommand{\arraystretch}{1.05} 
\caption{Distribution of DLT-enhanced AI studies across AI layers.}
\begin{tabular}{@{} l c c p{10cm} @{}}
\toprule
\makecell{\textbf{Enhanced}\\\textbf{AI Layer}} &
\makecell{\textbf{No. of}\\\textbf{Studies}} &
\textbf{(\%)} &
\textbf{Application Domain} \\
\midrule
Data           & 9 & \cellcolor{g4}{33.3} & Data provenance; secure \acs{ai} data pipelines; \acs{fl} data sharing; \acs{aigc} provenance \\
Model          & 7 & \cellcolor{g3}{25.9} & Federated learning coordination/aggregation; decentralised model evaluation \\
Application    & 5 & \cellcolor{o1}{18.5} & Decentralised \acs{ai} marketplaces; multi-agent coordination \\
Inference      & 4 & \cellcolor{o1}{14.8} & Inference verification and audit; explainability logging \\
Infrastructure & 2 & \cellcolor{r3}{7.4}  & Decentralised compute/training; hybrid on-/off-chain AI execution \\
\bottomrule
\end{tabular}
\label{tab:blockchain_enhanced_AI_dist}
\end{table*}

Across all studies, the most frequently addressed objectives were security (40.7\%), auditability (37.0\%), and trust (37.0\%). Privacy-related enhancements were reported in 18.5\% of publications, followed by incentive alignment (7.4\%), while explainability and interoperability were each addressed in 3.7\% of the reviewed works.

\begin{figure*}[t]
    \centering
    \includegraphics[width=1.9\columnwidth]{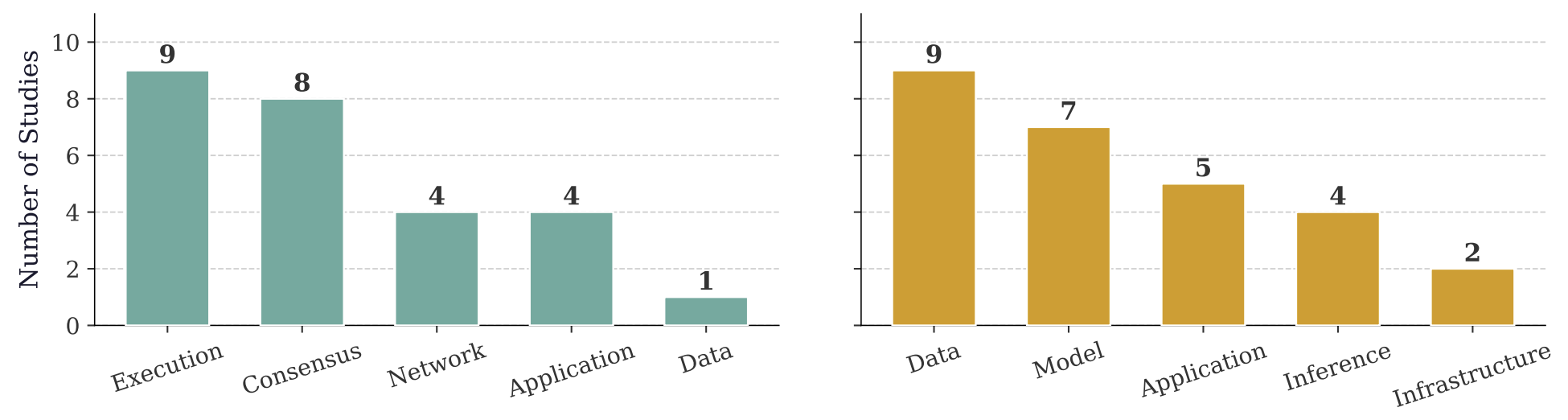}
    \caption{Distribution of reviewed studies by enhanced layer across the two directions of AI–DLT integration. The figure presents two complementary research directions: the left panel shows studies in which AI enhances DLT, while the right panel shows studies in which DLT enhances AI.}
    \label{fig:ai_chain}
\end{figure*}

Taken together, the layer distributions across both enhancement directions reveal distinct concentrations of research, as shown in \autoref{fig:ai_chain}.

\subsubsection{Mechanisms and Implementation Focus}

Studies used DLT execution mechanisms (e.g., smart contracts and programmable registries), consensus protocols, and privacy-preserving cryptography to improve AI systems.

\begin{itemize}

\item \textbf{Smart contracts and tokenisation:} 
Employed for model registration, provenance tracking, and incentive management, while NFTs represented ownership in AI marketplaces and AIGC provenance~\cite{battah2022blockchain, truong2024trust, fitriawijaya2024integrating}.

\item \textbf{Consensus and on-chain validation:} 
Novel consensus schemes such as \acf{poi} and \acf{potc} supported model aggregation and inference~\cite{guo2025verifiable, bi2022iot, muhammad2025blockchain}, while multi-agent frameworks embedded consensus mechanisms to prevent collusion among participating agents~\cite{chen2024blockagents}.

\item \textbf{Privacy-preserving mechanisms:} Studies combined \acf{dp}, \acf{smpc}, \acf{he}, and \acs{zkp}s with \acs{tee}s (\acs{sgx}, TrustZone) for privacy and verifiable execution \cite{kr2025blockchain, bhardwaj2025explainable, paulin2023histotrust, park2024brain}.

\item \textbf{Off-chain storage and hybrid architectures:} \acf{ipfs} and cross-chain setups supported scalable federated learning and provenance \cite{kang2022communication, kumar2023blockchain, witanto2022toward}.

\item \textbf{Identity and reputation:} Decentralised identifiers (\acs{did}s), verifiable credentials (\acs{vc}s), and reputation systems manage identity and participation in federated and agent-based AI systems \cite{papadopoulos2021privacy, mhamdi2022trust, truong2024trust}.

\end{itemize}

Implementations mainly used Ethereum, Hyperledger Fabric, or consortium DLTs. Off-chain storage and APIs managed model data, and one study demonstrated a public testnet deployment for decentralised AI training \cite{wang2025aiarena}. No study reported production-scale deployment.

\section{Discussion}
\label{sec:discussion}

\subsection{Key Insights}
\label{subsec:insights}
In this review, results show that AI is increasingly used to enhance DLT functionality, with most efforts directed toward improving security and performance. More than half of the studies focus on security auditing, anomaly detection, and vulnerability analysis, concentrated primarily in the Execution layer for smart-contract analysis and the Network layer for anomaly and fraud detection, where deep neural models report improved detection performance in contract auditing and runtime protection \cite{tang2023deep, deng2023smart, liu2023smart}. Reinforcement learning dominates at the Consensus layer, where agents tune protocol parameters and address malicious behaviour to improve throughput and reliability in permissioned DLTs \cite{li2023auto, goh2022secure, villegas2025adaptive}. These findings indicate a shift from static DLT operations toward adaptive, learning-driven systems based on network conditions and attack patterns. This evolution suggests that AI techniques align with DLT’s needs for adaptive consensus, network-aware resource control, and resilient security governance.

A second trend is the emergence of large and generative models within DLT execution and application environments, particularly for smart-contract auditing and development, and oracle reasoning. Studies from 2024–2025 introduced LLMs for explainable smart-contract auditing \cite{chen2025cryptic, yu2025smart}, semantically grounded oracle reasoning \cite{zeng2025connecting}, and autonomous smart-contract development \cite{khalid2025evaluating}. These works extend applications beyond control and detection toward semantic understanding and automation, where LLMs are used to analyse and generate DLT-related logic. Meanwhile, creative frameworks such as PoLe~\cite{lan2021proof} and PoUW~\cite{chenli2021dlchain} repurpose mining energy into machine learning computation, indicating overlap between distributed AI training and DLT consensus. Across all categories, experimentation remains largely confined to prototypes and emulations (e.g., Hyperledger Fabric, Ethereum testnets), but reported performance metrics: throughput, F1-scores, and explainability indices, suggest improvements over non-AI baselines, though evaluations are typically conducted in controlled settings of AI-driven DLT optimisation.

Conversely, DLT is often used as a trust and coordination substrate for distributed AI. The main application area is federated learning in privacy-sensitive domains such as healthcare and IoT, where DLT provides an immutable audit trail and coordination mechanism for model updates \cite{guo2025verifiable, kr2025blockchain}. 
To address the incentive problem directly, several works design consensus rules around the training task itself. Proof of Contribution and Proof of Trust are notable examples of this approach, where the validity of a block is conditional on the quality of the submitted model update \cite{bi2022iot, bhardwaj2025explainable}. Beyond training coordination, roughly one-third of the reviewed studies focused on data integrity and provenance, logging dataset origins, preprocessing steps, or inference outputs on permissioned DLTs for auditability and accountability \cite{soldatos2021blockchain, witanto2022toward, olawale2024cybersecurity}. This reflects a shift from optimizing accuracy toward ensuring traceability, explainability, and ethical data use.

Another trend is the combination of DLT and privacy-preserving AI through cryptographic integration. Transparent ledgers are augmented with differential privacy, homomorphic encryption, secure multi-party computation, and zero-knowledge proofs to protect sensitive gradients or records \cite{guo2025verifiable, kr2025blockchain, bhardwaj2025explainable}. Federated and healthcare-oriented systems increasingly adopt such hybrids to balance openness and confidentiality while keeping heavy computation off-chain for efficiency. Simultaneously, tokenised AI ecosystems are emerging: smart contracts and NFTs are used to represent model ownership and reward contributions in decentralised marketplaces \cite{tulkinbekov2025doctrina, battah2022blockchain, truong2024trust}. Platforms such as Doctrina and AIArena illustrate incentive-based collaboration by rewarding nodes for verified training or inference outputs \cite{wang2025aiarena}. Beyond FL, DLT-enhanced AI is expanding to multi-agent coordination and generative AI provenance, enabling verifiable agent interactions and authenticated AI-generated content \cite{chen2024blockagents, fitriawijaya2024integrating}.

\subsection{Research Gaps and Challenges}

Despite these advances, AI-enhanced DLT remains largely pre-deployment and fragmented. No study reports production-scale validation, underscoring persistent integration and trust challenges. Determinism, verifiability, and explainability remain open issues: decentralised nodes require reproducible results, adaptive models can yield divergent inferences, and reproducible verification benchmarks are still missing. Some studies mitigate this using oracle-level commit–reveal verification or rule-based validation \cite{liu2023smart, zeng2025connecting}, but no general framework for on-chain AI consensus currently exists. Furthermore, the research landscape is uneven: Execution, Application, and Consensus layers together account for 80.8\% of studies, while Network and especially Data remain comparatively underexplored. Adversarial robustness of AI models is rarely evaluated, even though DLT’s open environments expose these algorithms to data poisoning and manipulation risks. Energy-efficient and privacy-preserving deployment of AI systems and agents, particularly in edge and decentralised environments, also remain largely theoretical.

Scalability, interoperability, and empirical validation continue to be major bottlenecks. Most studies rely on permissioned DLTs or simulated testnets, with only AIArena reporting a deployed test network at scale \cite{wang2025aiarena}. Public chains remain constrained by transaction cost and latency, while hybrid off-chain storage (e.g., IPFS) introduces new trust assumptions \cite{kumar2023blockchain,witanto2022toward}. Interoperability remains limited: implementations are fragmented across Ethereum, Hyperledger, and consortium DLTs with inconsistent identity and provenance standards \cite{papadopoulos2021privacy}. Moreover, current evaluations emphasize security and auditability but rarely measure end-to-end AI performance impacts, generalisation, or long-term robustness. This limits understanding of how DLT integration affects model efficiency or user trust in production settings.

\subsection{Future Directions and Outlook}

Future efforts should focus on verifiable AI pipelines for DLT integration. Promising directions include consensus-aware model verification, lightweight explainability mechanisms for on-chain inference, and combined rule-based and learning-based approaches. Expanding AI applications to the Network and Data layers, such as congestion control, routing optimisation, learned storage, and state/indexing efficiency, could help balance the current bias toward Execution- and Consensus-layer enhancement. Cross-layer designs that integrate useful-work consensus, edge-aware AI scheduling, and federated DLT learning may support more scalable and energy-efficient systems. Collaborative testnets and reproducible cross-layer benchmarks integrating both AI and DLT components could support the transition from prototypes to deployed systems.

Further research should also address scalability, interoperability, and accountability. Layer-2 and cross-chain solutions may help address throughput bottlenecks, while unified schemas for AI provenance, digital identity, and smart-contract templates would enhance interoperability \cite{kang2022communication} and support emerging forms of model governance. Embedding explainability more deeply into DLT and AI workflows, for example through LLM-generated analyses and on-chain audit trails, may improve transparency and human oversight, particularly in autonomous decision systems. Empirical validation beyond controlled testbeds remains the critical missing step, particularly in sectors where auditability requirements and data sensitivity make DLT integration most justifiable on practical grounds. Additionally, DLT-based AI governance remains underexplored: on-chain mechanisms could support community-driven model tuning, audit explainability records, or reward fairness, accountability, and energy-efficient behaviour \cite{muhammad2025blockchain}. Extending the focus from trust and privacy toward alignment, sustainability, and fairness would position DLT not only as a verifier of AI but also as a participant in shaping its ethical and societal impact, suggesting potential extensions toward supporting AI reasoning and verification.

\section{Conclusion}

In this work, we analysed recent developments in the integration of \acs{ai} and DLT, synthesising findings from $53$ peer-reviewed studies published between 2020 and 2025. Using layered reference models, each composed of a five-layer architecture, the work categorised how \acs{ai} enhances DLT at the data, network, consensus, execution, and application layers, and how DLT enhances \acs{ai} across the infrastructure, data, model, inference, and application layers, enabling structured comparison across both enhancement directions. In summary, we investigated the synergy between the two technologies at different architectural layers.
The analysis highlights recent developments in \acs{ai}-driven DLT enhancement, including advances in consensus tuning, security auditing, anomaly detection, and autonomous contract analysis. \acs{rl}, \acs{dl}, and \acs{llm}s were frequently applied to improve DLT performance, programmability, and security. Conversely, DLT supports \acs{ai} through verifiable data provenance, \acs{fl} coordination, privacy-preserving computation, and decentralised \acs{ai} marketplaces. Smart contracts, distributed consensus, and cryptographic techniques are used to support trust and transparency in \acs{ai} workflows.
Across both directions, the findings reveal theoretical and practical advances, and highlight open areas such as verifiable \acs{ai} execution, provenance standards, decentralised learning, privacy-preserving architectures, cross-chain interoperability, and \acs{ai}-enabled consensus mechanisms. Overall, \acs{ai} and DLT address different system requirements, with \acs{ai} improving adaptability and DLT supporting transparency, coordination, and traceability.
These emerging research streams suggest that the convergence of \acs{ai} and DLT may contribute to more trustworthy and accountable systems.
Future research should focus on expanding the proposed framework beyond academic prototypes toward real-world deployments and cross-domain validation. This includes applying the layered model to operational systems, evaluating scalability in real environments, and examining interoperability with existing digital infrastructures.
Broader empirical studies across sectors and platforms would help assess practical constraints, reveal new design requirements, and inform real-world adoption.
The conclusions drawn here are based on a careful reading of the existing literature, although the study necessarily remains selective. The works considered provide a useful basis for outlining the main directions of a rapidly evolving field. A broader analysis of additional contributions would help refine this overview.

\section*{ACRONYMS}
\addcontentsline{toc}{section}{ACRONYMS}
\printacronyms[heading=none]

\bibliographystyle{IEEEtran}
\bibliography{ref}

\end{document}